\documentclass{article}
\usepackage{spconf,amsmath,graphicx}
\usepackage{booktabs} 
\usepackage{graphics}
\usepackage{adjustbox}
\usepackage{graphicx}
\usepackage{tabularx}
\usepackage{comment}
\usepackage{subfig}
\usepackage[justification=centering]{caption}
\usepackage{amsmath}
\usepackage{cite}
\usepackage{tikz}
\usepackage{amsmath,amssymb,amsfonts}
\usepackage{algorithmic}
\usepackage{graphicx}
\usepackage{booktabs} 
\usepackage{multirow}
\usepackage{svg}
\usepackage{multicol}
\usepackage{comment} 
\usepackage[utf8]{inputenc} 
\usepackage{textcomp}
\usepackage{xcolor}

\title{Structure-aware Audio-to-Score Alignment using progressively dilated Convolutional Neural Networks}
\name{Ruchit Agrawal $^{\dagger}$ \qquad Daniel Wolff $^{\star}$ \qquad Simon Dixon $^{\dagger}$ 
      \thanks{This research is supported by the European Union’s Horizon 2020 research and innovation programme under the Marie Skłodowska-Curie grant agreement number 765068.}}
  \address{$^{\dagger}$ Centre for Digital Music, Queen Mary University of London, UK \\
      $^{\star}$ Institute for Research and Coordination in Acoustics/Music, Paris, France}
\begin{document}
\ninept
\maketitle
  \tikz [remember picture, overlay] %
     \node [shift={(50mm,22mm)}] at (current page.south west) %
     [anchor=south west] %
     {\includegraphics[width=6mm]{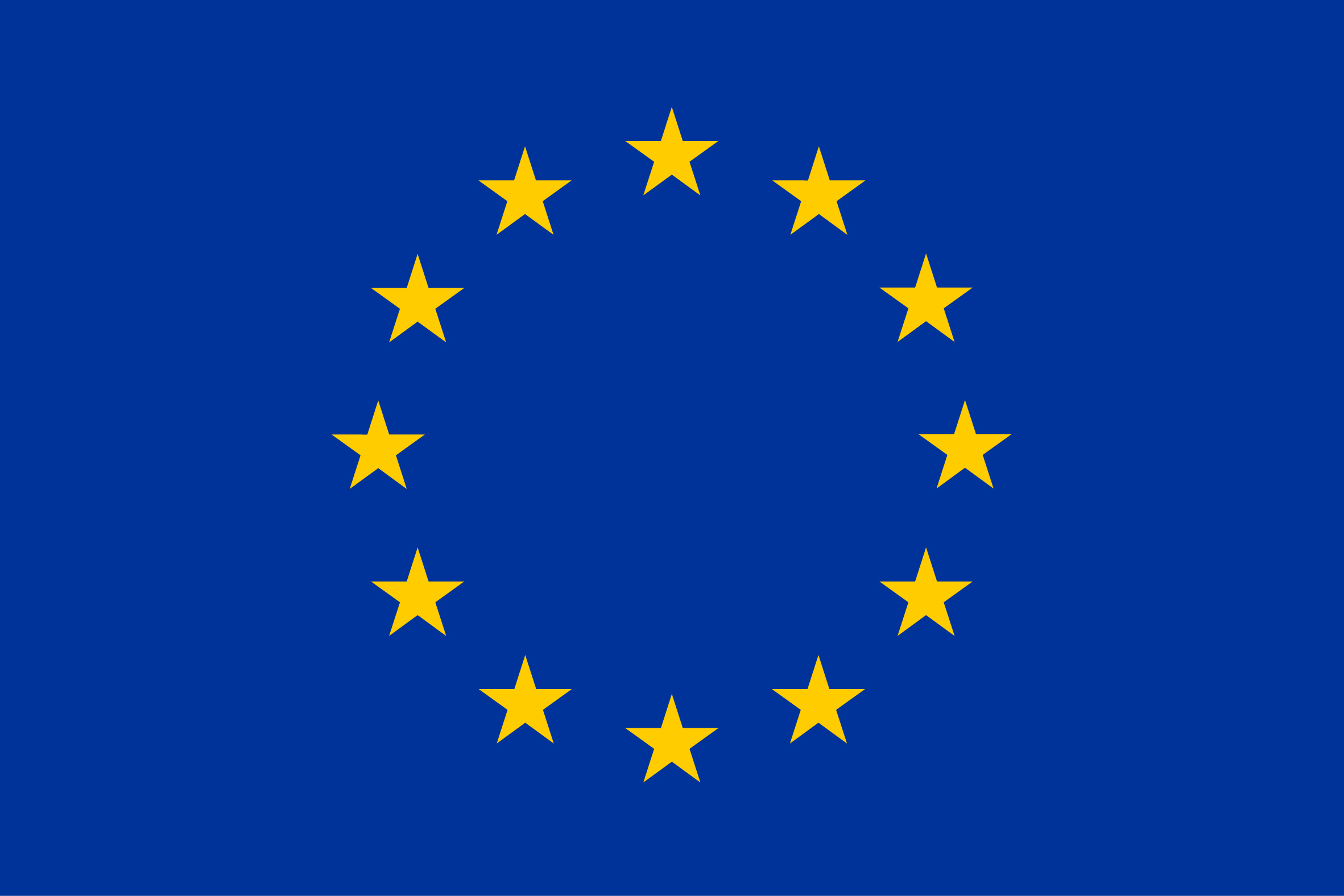}};
\vspace{-0.5cm}
\begin{abstract}
The identification of structural differences between a music performance and the score is a challenging yet integral step of audio-to-score alignment,  an important subtask of music information retrieval. We present a novel method to detect such differences between the score and performance for a given piece of music using progressively dilated convolutional neural networks. Our method incorporates varying dilation rates at different layers to capture both short-term and long-term context, and can be employed successfully in the presence of limited annotated data. We conduct experiments on audio recordings of real performances that differ structurally from the score, and our results demonstrate that our models outperform standard methods for structure-aware audio-to-score alignment.
\end{abstract}
\vspace{-0.2cm}
\section{Introduction}
\vspace{-0.2cm}
The analysis of 
music performance is a challenging area of Music Information Retrieval (MIR), owing to multiple factors such as suboptimal recording conditions, structural differences from the score, and subjective interpretations. 
Deviations from the structure and/or tempo prescribed by the score are 
common in several genres of music, particularly classical music and jazz \cite{widmer2016getting}.
Identification of such structural differences from the score is a challenging yet important step of performance analysis, especially for the audio-to-score alignment task. 
The alignment task aims at generating an accurate mapping between a performance audio and the score for a given piece of music.
The particular path through the score that the musician is going to take is difficult to predict, and although there have been several methods proposed in the recent past to handle these impromptu changes;  it still remains a problem that is not fully solved \cite{arzt2016flexible}.
\vspace{0.1cm}
 \par The majority of approaches for audio-to-score alignment are based on Dynamic Time Warping (DTW) \cite{dixon2005line} or Hidden Markov Models (HMM) \cite{muller2015fundamentals}. These methods typically assume that the musician follows the score from the beginning to the end without inducing any structural modifications, which is often not the case in real world scenarios. The alignments computed using DTW-based methods are constrained to progress monotonically through the piece, and are thereby unable to model structural deviations such as repeats and jumps. Previous methods for handling structural changes during alignment are either reliant on robust Optical Music Recognition (OMR) to detect repeat and jump directives \cite{Fremerey2010handling}; or on frameworks fundamentally different from DTW, such as the Needleman Wunsch Time Warping method \cite{grachten2013automatic}. The former method, called \begin{math}\textit{JumpDTW}\end{math}, requires manually annotated frame positions for block boundaries, initially provided by an OMR system, and is unable to model deviations that are not foreseeable from the score. The latter is unable to align repeated segments, since it does not introduce backward jumps, and is based on a waiting mechanism upon mismatch of the two streams. 
 Since structural changes like repeats and jumps can be arbitrarily added during a performance, especially during rehearsals; these are challenging to model using rule-based approaches, and machine learning methods offer promise at effectively addressing these challenges. 
 \par This paper is aimed at handling structural changes between the performance and the score for the offline audio-to-score alignment task. We propose a custom Convolutional Neural Network (CNN) based architecture for modeling these differences, coupled with a flexible DTW framework to generate the fine alignments. 
 Our method does not require a large corpus of hand-annotated data and can also be trained exclusively using synthetic data if hand-annotated data is unavailable.
Our architecture employs progressively dilated convolutions in addition to standard convolutions, with varying amounts of dilation applied at different layers of the network. 
The primary motivation behind our architecture is that it allows us to effectively capture both short-term and long-term context, 
using much fewer parameters than sequential models such as recurrent neural networks, and without facing the vanishing gradient problem. We conduct experiments on piano music and compare our method with three major alignment approaches; namely \begin{math}\textit{JumpDTW}\end{math} \cite{Fremerey2010handling}, \begin{math}\textit{NWTW}\end{math} \cite{grachten2013automatic}, and \begin{math}\textit{MATCH}\end{math} \cite{dixon2005line}. 
We demonstrate results on two different test sets containing real performances. Our method yields higher performance than previous methods proposed for handling structural deviations, without requiring manually annotated data; and noticeably outperforms these methods given a limited amount of annotated data.  To the authors' knowledge, this is the first method to employ dilated convolutional neural networks for structure-aware audio-to-score alignment. 
\begin{figure*}[th]
  \vspace{-2.6cm}
  \centering
  \includegraphics[width=6.5in]{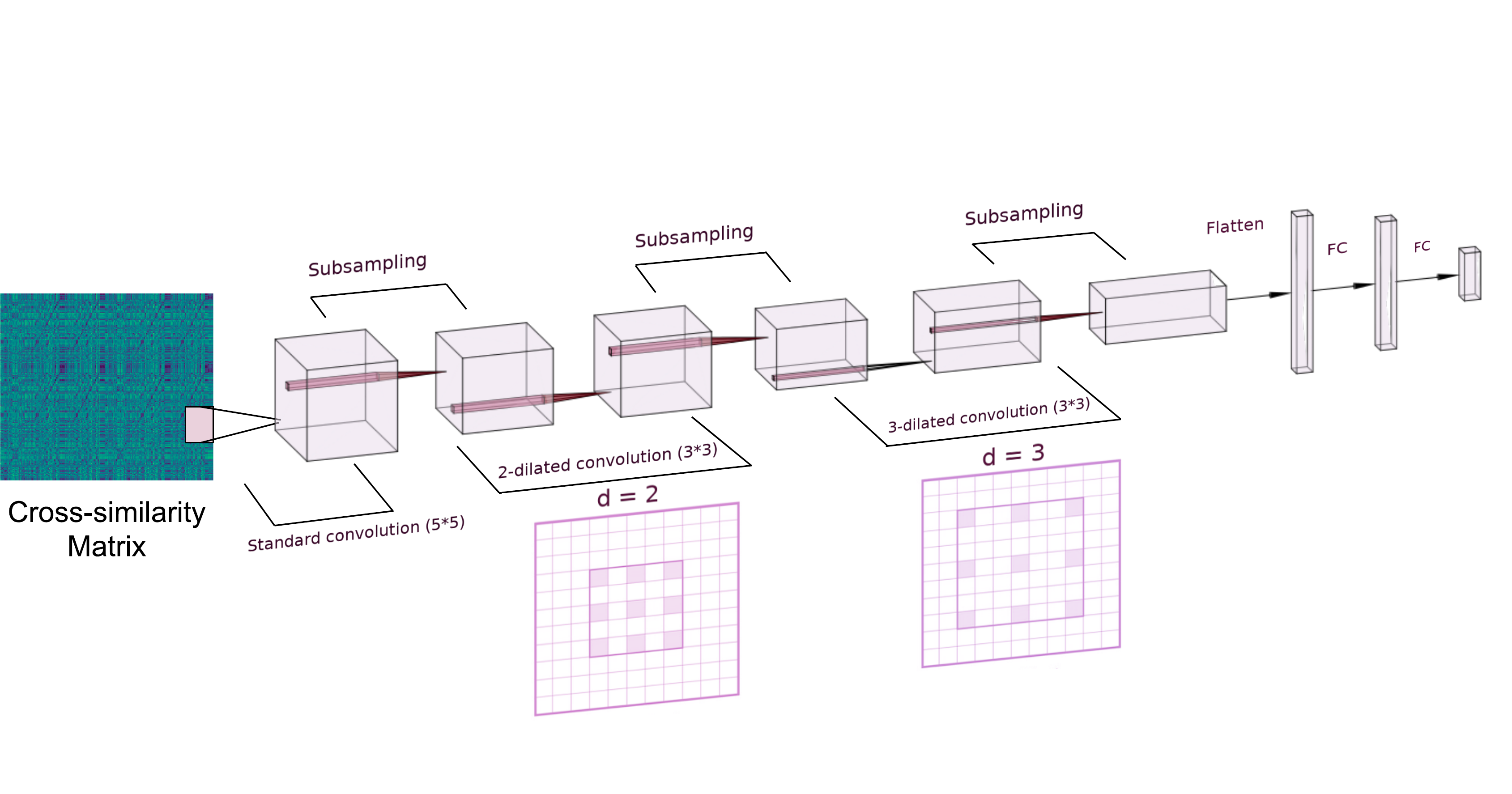}
  \vspace{-0.7cm}
  \caption{Schematic diagram illustrating the general architecture of our models.\\ \textit{d: Dilation rate, FC: Fully connected layer}}
  \vspace{-0.4cm}
  \label{fig:pipeline}
\end{figure*}
\vspace{-0.3cm}
\section{Related Work}\label{related}
\vspace{-0.2cm}
Early work on structure-aware MIR focuses on structural segmentation of musical audio by constrained clustering \cite{levy2008structural} and music repetition detection using histogram matching \cite{tian2009histogram}. 
Arzt and Widmer \cite{arzt2010towards} propose a multilevel matching and tracking algorithm to deal with issues in score following due to deviations from the score in live performance. 
A challenge faced by this approach appears when complex piano music is played with a lot of expressive freedom in terms of tempo changes. Hence, they propose methods to estimate the current tempo of a performance, which could then be used to improve online alignment  \cite{arzt2010simple}. This is similar to the work proposed by Müller et al. \cite{muller2009towards}, wherein they develop a method for automatic extraction of tempo curves from music recordings
by comparing performances with neutral reference representations.

\par Work specifically on structure-aware audio-to-score alignment includes \begin{math}\textit{JumpDTW}\end{math} \cite{Fremerey2010handling} and Needleman-Wunsch Time Warping (\begin{math}\textit{NWTW}\end{math}) \cite{grachten2013automatic}  among others. 
Fremerey et al. \cite{Fremerey2010handling} focus on tackling structural differences induced specifically by repeats and jumps, using a novel DTW variation called \begin{math}\textit{JumpDTW}\end{math}. This method identifies the ``block sequence" taken by a performer along the score, however it requires manually annotated block boundaries to yield robust performance, which are generally not readily available at test time for real world applications. Additionally, it cannot align deviations that are not foreseeable from the score. \begin{math}\textit{NWTW}\end{math} \cite{grachten2013automatic}, on the other hand, is a pure dynamic programming method to align
music recordings that contain structural differences. This method is an extension of the classic Needleman-Wunsch sequence alignment algorithm  \cite{likic2008needleman}, with added capabilities to deal with the time warping aspects of aligning music performances. A caveat of this method is that 
it does not successfully align repeated segments owing to its waiting mechanism, which skips unmatchable parts of either sequence, and makes a clean jump when the two streams match again. 
\vspace{-0.1cm}
 \par Apart from \begin{math}\textit{JumpDTW}\end{math} and \begin{math}\textit{NWTW}\end{math}, which focus on offline alignment, work on online score following \cite{nakamura2015real} has demonstrated the effectiveness of HMMs for modeling variations from the score. While this method focuses on real-time score following for monophonic music, our work deals with offline audio-to-score alignment for polyphonic music performance. 
Another work related to ours is that proposed by Jiang et al.\cite{jiang2019offline}, which  focuses on offline score alignment for the practice scenario. Similar to Nakamura et al.\cite{nakamura2015real}, their approach is also based on HMMs; but they propose using pitch trees and beam search to model skips. However, their method struggles with pieces containing both backward and forward jumps, which is an important challenge we tackle using our progressively dilated convolutional models.
Recent work on audio-to-score alignment has demonstrated the efficacy of multimodal embeddings \cite{dorfer2018learning}, reinforcement learning \cite{dorfer2018learning2}, \cite{henkel2019score} and learnt frame similarities \cite{agrawal2021learning}, albeit these are not structure-aware methods. Very recently, Shan et al. propose Hierarchical-DTW \cite{shan2020improved} to automatically generate piano score following videos given an audio and a raw image of sheet music. Their method is reliant on an automatic music transcription system \cite{hawthorne2017onsets} and a pre-trained model to extract bootleg score representations \cite{tanprasert2020midi}. It struggles when this representation is not accurate, and also on short pieces containing jumps. While they work with raw images of sheet music and generate score following videos; our method works with symbolic scores, is not reliant on other pre-trained models, and performs well on both short and long pieces. 
\begin{table*}[th]
  \vspace*{-0.6cm}
  \begin{adjustbox}{max width=\textwidth}
\begin{tabular}{ccccccccccccc}
\toprule
\hline 
\multirow{2}{*}{\textbf{Model}} & 
\multicolumn{4}{c}{\textit{On Mazurka dataset}}  
& \multicolumn{4}{c}{\textit{With structural differences (Tido)}} & \multicolumn{4}{c}{\textit{Without structural differences (Tido)}}
\tabularnewline
  & \textbf{$<$25ms}& \textbf{$<$50ms} & \textbf{$<$100ms} & \textbf{$<$200ms} & \textbf{$<$25ms} &\textbf{$<$50ms} & \textbf{$<$100ms} & \textbf{$<$200ms}
  &
  \textbf{$<$25ms} &
  \textbf{$<$50ms} & \textbf{$<$100ms} & \textbf{$<$200ms}
  \\
\midrule 
 \begin{math}\textit{MATCH}\end{math}\cite{dixon2005line} & 64.8 & 72.1 & 77.6 & 83.7 & 61.5 & 70.4 & 74.6 & 80.7 & 70.2 & 78.4 & 84.7 & 90.3  \\
\midrule
 \begin{math}\textit{JumpDTW}\end{math} \cite{Fremerey2010handling} & 65.8 & 75.2 & 79.8 & 85.7 & 69.1 & 77.2 & 82.0 & 88.4 & 68.7 & 77.5 & 82.1 & 88.9  \\
\midrule 
 \begin{math}\textit{NWTW}\end{math} \cite{grachten2013automatic} & 67.6 & 75.5 & 80.1 & 86.2 & 68.6 & 75.8 & 80.7 & 87.5 & 68.4 & 77.1 & 82.8 & 89.4  \\
\midrule 
     \emph{CNN}${}_{1+1}$ & 68.2 & 75.7 & 80.5 & 87.1   & 70.4 & 78.3 & 83.4 & 90.1 & 69.3 & 78.0 & 84.1 & 89.3 \\
\midrule 
      \emph{DCNN}${}_{2+2}$ & \textbf{69.9} & 76.4 & 81.6 & 88.9 & 72.7 & 80.1  & 84.5 & 91.4 & \textbf{71.4} & 79.5 & 85.3 & 90.5 \\
\midrule 
    \emph{DCNN}${}_{2+3}$ & 69.7 & \textbf{77.2} & \textbf{82.4} & \textbf{89.8} & \textbf{73.9} & \textbf{81.3} & \textbf{85.6} & \textbf{92.8} & 71.0 & \textbf{80.3} & \textbf{85.8} & \textbf{91.8} \\
\midrule 
     \emph{DCNN}${}_{3+3}$ & 69.2 & 76.1 & 81.2 & 88.7  & 72.3 & 79.5 & 84.2 & 90.4  & 70.6 & 78.8 & 84.9 & 91.2\\
  \midrule
  \emph{DCNNsyn}${}_{2+3}$ & 68.1 & 75.9 & 80.7 & 87.5  & 70.5 & 78.6 & 83.8 & 90.5 & 69.2 & 78.3 & 84.6 & 89.8  \\
\midrule 
\bottomrule
\end{tabular}
\end{adjustbox}
\vspace{0.1cm}
\caption{Alignment accuracy in \% on  the Mazurka and Tido test sets \\ \emph{DCNN}${}_{m+n}$: Dilated CNN model with dilation rates of $m$ and $n$ at the second and third layer respectively \\
}
\vspace{-0.5cm}
\label{results2}
\end{table*}
\vspace{-1.0cm}
\section{Proposed
Method}\label{method}
\vspace{-0.2cm}
We present a novel method to detect structural differences for audio-to-score alignment using a custom convolutional architecture.  The closest to our method is the work proposed by  Ullrich et al.\cite{ullrich2014boundary}, an adaptation of the onset detection method proposed by Schl{\"u}ter et al.\cite{schluter2014improved}.
While Ullrich et al.\cite{ullrich2014boundary} train a CNN as a binary classifier on spectrogram excerpts for popular music (using hand annotated data) to detect boundaries in a song, we focus on structure-aware alignment of a performance to the score, with the cross-similarity matrices as inputs. Additionally, we incorporate dilation in our CNN models to incorporate multi-scale context. 
\par Our architecture combines standard convolution and dilated convolution \cite{yu2015multi}, with varying amounts of dilation applied at different layers of the network. 
We approach the sturctural deviation identification problem as a multi-label prediction task, and train our models to detect synchronous subpaths between the score and the performance. These subpaths are detected by means of inflection points, which encode the positions of structural differences between the two streams. 
Figure \ref{fig:pipeline} illustrates the general architecture of our models. Our networks operate on the cross-similarity matrix between the score and performance and predict the ($x$, $y$) co-ordinates of the inflection points as the output.
We employ dilated convolution at the second and the third layer, and standard convolution at the first layer.
The dilated convolution operation \cite{yu2015multi} of a discrete function $F$ with a discrete filter $f$ on an element $\textbf{p}$ is defined as follows:
\vspace{-0.1cm}
\begin{equation}
(F*_df)(\textbf{p}) = \sum\limits_{\textbf{t}}F(\textbf{p}-d\textbf{t})f(\textbf{t})
\end{equation}
where $d$ is the factor by which the kernel is inflated, referred to as the dilation rate. Inflating the kernel using dilation allows us to incorporate larger context without increasing the number of parameters, which is essential for modeling structure. The receptive field is exponentially increased, and for a dilation rate $d$, a kernel of size $m$ effectively works as a kernel of size $m'$  as follows:
\vspace{-0.1cm}
\begin{equation} m' = m + (d-1)*(m-1) 
\end{equation} This facilitates the incorporation of context better than standard convolutions, which can only offer linear growth of the effective receptive field as we move deeper into the network. We conduct experiments using varying dilation rates at different layers of the network to determine the optimal amount of dilation at each layer. The motivation behind using varying amounts of dilation is to incorporate both short-term and long-term context to model structure, 
which has proven to be useful in computer vision as well as natural language processing tasks \cite{lee2017going, agrawal2018contextual}.
We observe that progressively increasing dilation as we move deeper into the network produces the best results for detecting the structural differences. We compare the results of our networks with previous methods proposed for handling structural changes during alignment (\cite{Fremerey2010handling}, \cite{grachten2013automatic}) as well as a baseline CNN model trained without any dilation.  
\vspace{0.1cm}
 \par In order to generate the fine alignments, the inflection points predicted by our dilated convolutional models are employed as potential jump positions to assist a DTW-based alignment algorithm. We implement such an extended DTW framework, inspired by \begin{math}\textit{JumpDTW}\end{math}  \cite{Fremerey2010handling}, to allow for jumps between the synchronous subpaths.  
We assume $X$= $(x_1, x_2,..., x_p)$ to be the feature sequence corresponding to the performance and $Y$ = $(y_1, y_2,..., y_q)$ to be the feature sequence corresponding to the score. Furthermore, let ($a_i$, $b_i$) denote the ($x$, $y$) co-ordinates of the $i_{th}$ inflection point, and $N$ denote the total number of inflection points. The odd numbered inflection points correspond to the end of the synchronous subpaths and the even numbered points correspond to the beginning of the subpaths. We modify the classical DTW framework to extend the set of possible predecessor cells for the cell $(a_i, b_i)$ for all \begin{math} i \in \{2, 4, 6, .., N\} \end{math}, as follows:
\vspace{-0.2cm}
\begin{equation}
D(m, n)  = e(m, n) + min\begin{cases}
D(m, n-1) \\ D(m-1, n) \\  D(m-1,  n-1) \\
D(a_{i-1}, b_{i-1}) \hspace{0.1cm} \forall (m, n) = (a_i, b_i),\\ 
\hspace{2.2cm} i \in \{2, 4, ..., N\}
\end{cases}
\end{equation}
where $e(m, n)$ is the Euclidean distance between points $x_m$ and $y_n$, and $D(m, n)$ is the total cost to be minimized for the path until the cell $(m, n)$. The path which yields the minimum value for $D(p, q)$ is taken to be the optimal alignment path between the performance and score sequences.

\begin{figure*}%
    \centering
   \vspace{-1.0cm}
  {{\includegraphics[width=2.9cm, height=2.7cm]{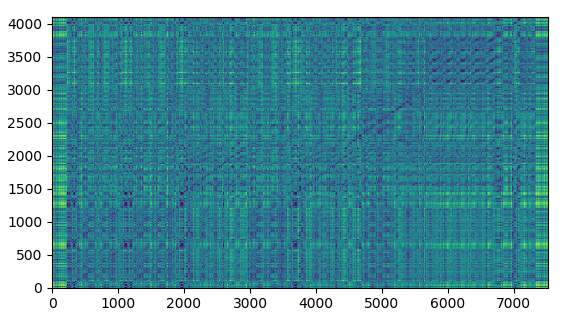} }}%
    \qquad
    {{\includegraphics[width=2.9cm, height=2.7cm]{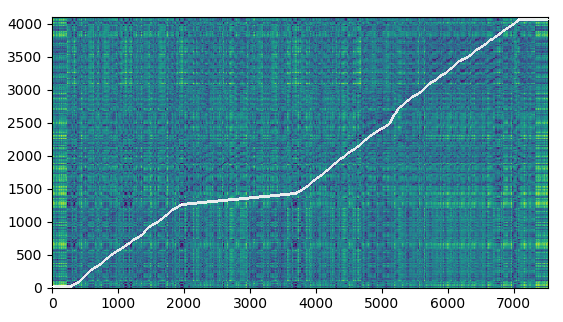} }}%
    \qquad
    {{\includegraphics[width=2.9cm, height=2.7cm]{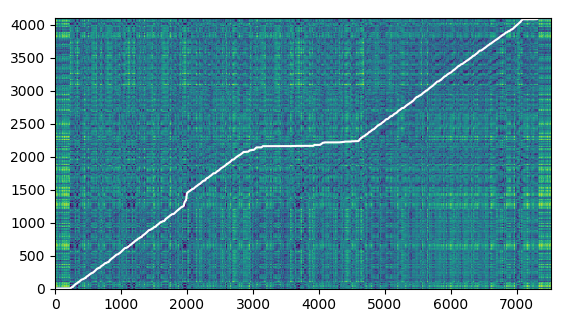} }}%
    \qquad
    {{\includegraphics[width=2.9cm, height=2.7cm]{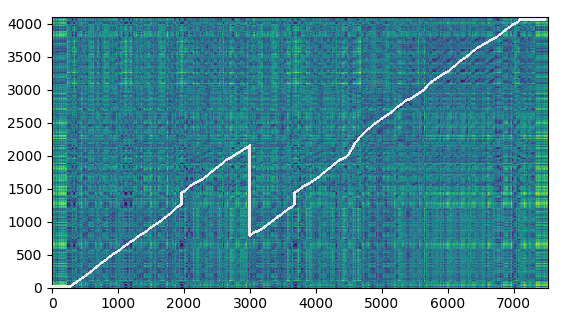} }}%
    \qquad
    {{\includegraphics[width=2.9cm, height=2.7cm]{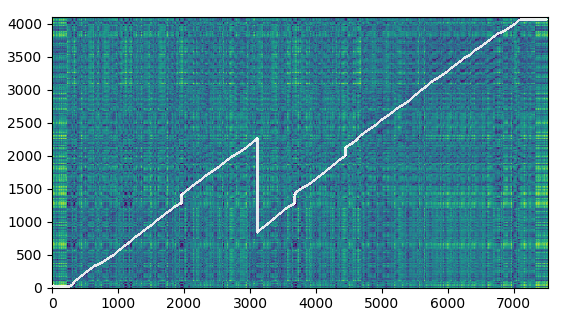} }}%
    \qquad
    
    {{\includegraphics[width=2.9cm, height=2.6cm]{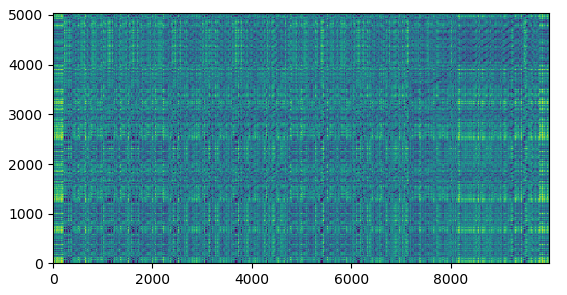} }}%
    \qquad
    {{\includegraphics[width=2.9cm, height=2.6cm]{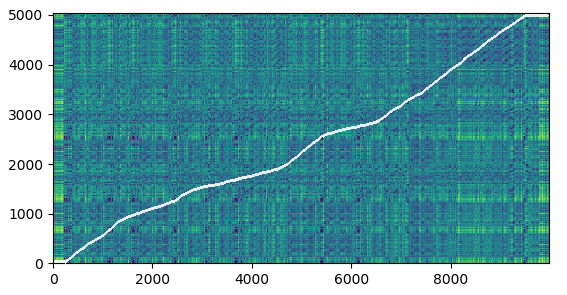} }}%
    \qquad
    {{\includegraphics[width=2.9cm, height=2.6cm]{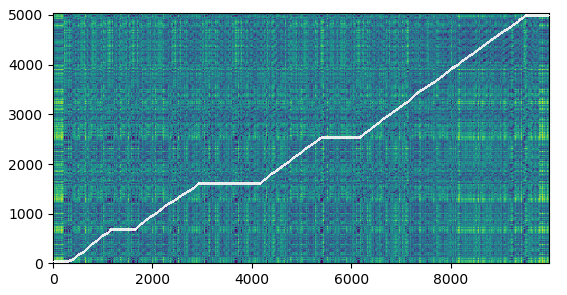} }}%
    \qquad
    {{\includegraphics[width=2.9cm, height=2.6cm]{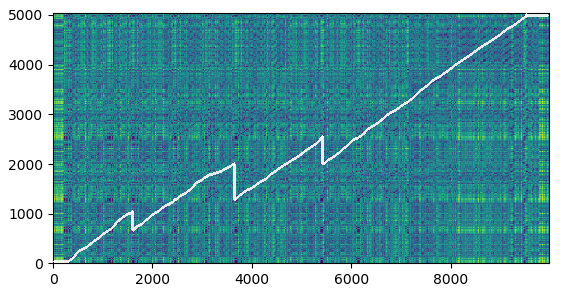} }}%
    \qquad
    {{\includegraphics[width=2.9cm, height=2.6cm]{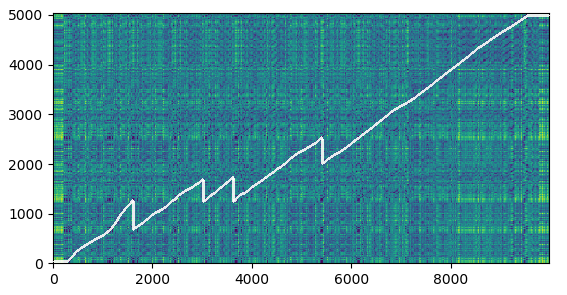} }}%
    \qquad
    \subfloat[Input]
    {{\includegraphics[width=2.9cm, height=2.8cm]{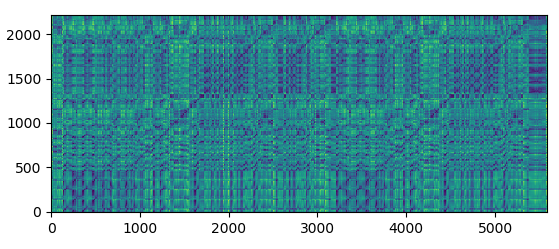} }}%
    \qquad
    \subfloat[\begin{math}\textit{JumpDTW}\end{math}]
    {{\includegraphics[width=2.9cm, height=2.8cm]{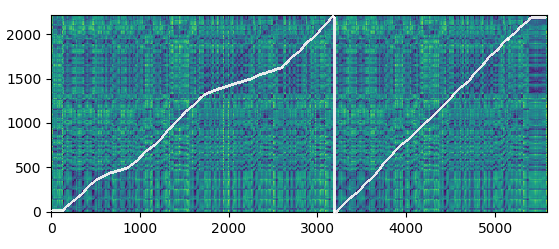} }}%
    \qquad
    \subfloat[\begin{math}\textit{NWTW}\end{math}]
    {{\includegraphics[width=2.9cm, height=2.8cm]{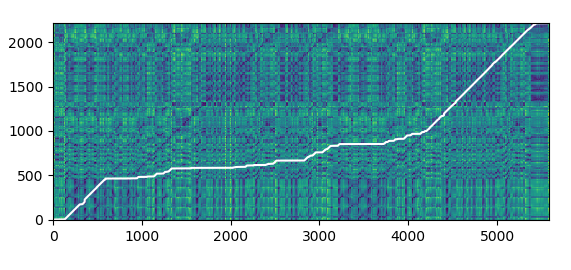} }}%
    \qquad
    \subfloat[\emph{DCNN}${}_{2+3}$]
    {{\includegraphics[width=2.9cm, height=2.8cm]{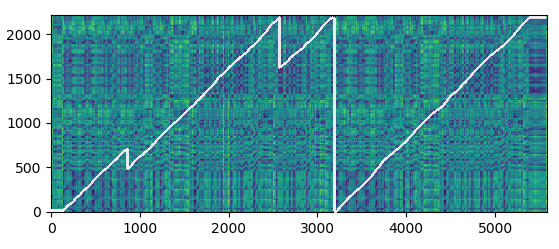} }}%
    \qquad
    \subfloat[Ground Truth]
    {{\includegraphics[width=2.9cm, height=2.8cm]{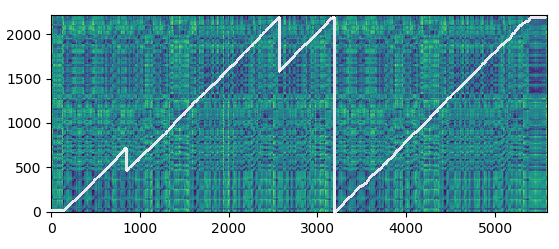} }}%
    \qquad
    
    \vspace{-0.1cm}
    \caption{Comparison of our alignment path with standard methods. \\Input: Cross-similarity matrix between score and performance, 
    X-axis: Frame index (performance), Y-axis: Frame index (score)}%
    \vspace{-0.5cm}
    \label{fig:comparison}%
\end{figure*}
\vspace{-0.3cm}
\section{Experiments and Results}\label{experiments}
\vspace{-0.2cm}
\subsection{Experimental Setup}\label{setup}
\vspace{-0.1cm}
We model the task of detecting the synchronous subpaths as a multi-label prediction task using progressively dilated CNNs, with each output label encoding a deviation in the performance from the score. 
A key challenge in modelling structural changes for alignment is the lack of hand annotated data, marked for repeats and jumps accurately at the frame level. This is also one of the caveats of \begin{math}\textit{JumpDTW}\end{math}, which is reliant on the accuracy of the OMR system to detect jump and repeat directives in the absense of manually annotated boundaries.
To overcome the lack of annotated training data, we generated synthetic samples containing jumps and repeats using the audio from the MSMD \cite{dorfer2018learning} dataset. 
Our training dataset contains 2625 pairs of audio recordings, corresponding to the MIDI score and performance respectively. 2475 of these are obtained from the MSMD dataset, with each piece utilized 4 times for varying number of repetitions (generated synthetically), and once without any repetition. In addition to synthetic data, we employed a small amount of hand annotated data from a private dataset procured from Tido UK Ltd., referred to as the Tido dataset further in the paper. The training set taken from the Tido dataset comprises audio pairs for 150 pieces, 80 of which contain structural differences. In addition to our models trained with different dilation rates on the entire training set, we also demonstrate the results obtained by our progressively dilated CNN model trained exclusively on synthetic data (\emph{DCNNsyn}${}_{2+3}$).
\par We compute the cross-similarity matrix for each performance-score pair using the Euclidean distance between the chromagrams corresponding to the score and performance respectively. We employ librosa\cite{mcfee2015librosa} to compute the chromagrams as well as the cross-similarity.
Our models consist of three convolutional and subsampling layers, with standard convolutions at the first layer and dilated convolutions with varying dilation rates at the second and third layer respectively. The output of the third convolutional and subsampling layer is sent through a flatten layer, following which it is passed through two fully connected layers of size 4096 and 1024 respectively to predict the (x, y) co-ordinates of the inflection points. The output of the final layer is a one-dimensional tensor of size 64, signifying that the model can predict up to 32 inflection points, with their $(x, y)$ co-ordinates in chronological order. The output is compared with the ground truth using the L2 regression loss, since we want to capture the distance of the predicted inflection points from the ground truth inflection points in time. The outputs of each layer are passed through rectified linear units to add non-linearity, followed by batch normalization before being passed as inputs to the next layer. We employ a dropout of 0.5 for the fully connected layers to avoid overfitting. Our batch size is 64 and the models are trained for 40 epochs, with early stopping. Our dilated CNN models are denoted as  \emph{DCNN}${}_{m+n}$, where $m$ and $n$ correspond to the dilation rates at the second and third layer respectively. 
\par We test the performance of our models on two different datasets, both containing recordings of real performances. We demonstrate the results of our models on the publicly available Mazurka dataset \cite{sapp2007comparative}. In order to analyze specific improvements for structurally different pieces, we also demonstrate results on subsets of the Tido dataset \emph{with} and \emph{without} structural differences. Both the subsets contain 75 pieces each. We compare the results obtained by our models with \begin{math} \textit{JumpDTW}\end{math} \cite{Fremerey2010handling}, \begin{math}\textit{NWTW}\end{math} \cite{grachten2013automatic}, \begin{math}\textit{MATCH}\end{math} \cite{dixon2005line} and a vanilla CNN model without dilation     \emph{CNN}${}_{1+1}$. The number of parameters of  our \emph{DCNN}${}_{m+n}$ networks is comparable with that of the baseline \emph{CNN}${}_{1+1}$ network.
For comparison with \begin{math}\textit{JumpDTW}\end{math}, we employ the SharpEye OMR engine to extract frame predictions for block boundaries from the  sheet images \cite{Fremerey2010handling}. These are then passed on to an implementation of \begin{math}\textit{JumpDTW}\end{math} to generate the alignment path. 
Similarly, we compare our models with an implementation of the \begin{math}\textit{NWTW}\end{math} method, and estimate the optimal gap penalty parameter $\gamma$  \cite{grachten2013automatic} on our data.
\vspace{-0.4cm}
\subsection{Results and Discussion}\label{results}
\vspace{-0.2cm}
The results obtained by our models are given in Table \ref{results2}. We report alignment accuracy in \%, 
where each value denotes the percentage of beats aligned correctly within the corresponding time durations of 25, 50, 100 and 200 ms respectively.
Our models show an increase of 2-5\% in alignment accuracy over \begin{math}\textit{JumpDTW}\end{math} and \begin{math}\textit{NWTW}\end{math} on the test subset containing structural differences and an increase of 1-3\% on the test subset not containing structural differences. Compared with \begin{math}\textit{MATCH}\end{math}, our models show an increase of 9-10\% on the subset with structural differences, and an increase of 1-2\% on the subset without structural differences. Overall accuracy on the Mazurka dataset suggests that our models perform better than \begin{math}\textit{MATCH}\end{math} by 4-6\% as well as the \begin{math}\textit{JumpDTW}\end{math} and \begin{math}\textit{NWTW}\end{math} frameworks by 1-4\% (Table 1, columns 1-4).
Our model trained exclusively on synthetic data (\emph{DCNNsyn}${}_{2+3}$) yields better alignment accuracy than \begin{math}\textit{JumpDTW}\end{math}, which requires manually labelled block boundaries to handle repeats and jumps \cite{Fremerey2010handling}. This emphasizes the applicability of our method in real-world scenarios with scarce availability of hand-annotated data. Our models noticeably outperform all methods when a limited amount of real data is added to the synthetic data during training (Table 1, rows 5-7).
\vspace{0.1cm}
\par The experimentation with different dilation rates reveals that progressively increasing dilation as we move deeper (\emph{DCNN}${}_{2+3}$) yields better results than models trained using equal amounts of dilation (\emph{DCNN}${}_{2+2}$, \emph{DCNN}${}_{3+3}$). Models trained with dilation at the first layer and those trained using dilation rates of 4 and higher did not yield improvement over the vanilla CNN model \emph{CNN}${}_{1+1}$ and hence are not reported. We speculate that progressively increasing dilation helps the model learn higher level features better further down the network. Manual inspection of the alignments confirmed that long-term context was better captured by the progressively dilated CNNs than other models, and they could detect larger deviations in addition to short ones. We demonstrate the alignment paths generated by the comparative methods for three performance-score pairs in Figure \ref{fig:comparison} to facilitate qualitative understanding of our results. 
While \begin{math}\textit{JumpDTW}\end{math} struggles with deviations from the score, \begin{math}\textit{NWTW}\end{math} struggles with cases containing both forward and backward jumps (Fig. \ref{fig:comparison}, example 1). This could be attributed to the waiting mechanism of \begin{math}NWTW\end{math}, which makes backward jumps especially challenging. Our model \emph{DCNN}${}_{2+3}$ struggles with cases where there are multiple deviations within a short time span (Example 2). We speculate that this is due to the larger receptive fields of the dilated convolutions, which, while capturing greater context, are sometimes unable to capture multiple inflection points within a small context. Our model is specifically able to model forward jumps better than both the methods  (Example 1), while also handling deviations not foreseeable from the score (Example 3).
\vspace{-0.1cm}
\par We demonstrate that progressively dilated convolutional neural networks are effective at detecting structural differences between the score and the performance for structure aware audio-to-score alignment. While we used chroma-based features for score-performance audio pairs, our method can also be used with raw or scanned images of sheet music using learnt features, for instance, using multimodal embeddings trained on audio and sheet image snippets \cite{dorfer2018learning}, \cite{balke2019learning}. Additionally, our method can also be utilized by frameworks other than DTW to generate the alignments. For instance, an unrolled score representation could be achieved via the inflection points predicted by our model, which could further be employed by score following approaches based on reinforcement learning \cite{dorfer2018learning2}, \cite{henkel2019score} for structure-aware tracking.
The advantage of our method is that it does not require manually labelled block boundaries, and can effectively deal with deviations from the structure given in the score, in both the forward and backward directions.
 In the future, we would like to experiment with parallel  dilation using different dilation rates and merging the learnt features. A current limitation of our method is the handling of trills and cadenzas, and we would also like to address these issues in future research. 
\bibliographystyle{IEEEbib}
\bibliography{refs}

\end{document}